\documentclass[twocolumn]{revtex4}

\usepackage{amsmath}
\usepackage{amsfonts}
\usepackage{amssymb}
\usepackage{bm}
\usepackage{tabularx}
\usepackage{graphicx}
\usepackage{color}
\usepackage{txfonts}

\def\RAMO{$R_{1-y}A_{y}$MnO$_3$}
\def\LPCMO{(La$_{5/8-x}$Pr$_{x}$)Ca$_{3/8}$MnO$_3$}
\def\LPSMO{(La$_{0.4}$Pr$_{0.6}$)$_{1.2}$Sr$_{1.8}$Mn$_2$O$_7$}
\def\EGSMO{(Eu$_{1-x}$Gd$_{x}$)$_{0.6}$Sr$_{0.4}$MnO$_3$}
\def\ESMO{Eu$_{0.6}$Sr$_{0.4}$MnO$_3$}
\def\L0{$L_0$}
\def\dLR{$\Delta L_\mathrm{R}$}
\def\dMR{$\Delta M_\mathrm{R}$}
\def\Mn3+{Mn$^{3+}$}

\begin{document}

\title{Dynamics of the lattice and spins in the phase-separated manganite \EGSMO}

\author{Haruka Taniguchi}
\email{tanig@iwate-u.ac.jp}
\author{Daichi Kimura}
\author{Michiaki Matsukawa}
\author{Tasuku Inomata}
\author{Satoru Kobayashi}
\affiliation{Department of Physical Science and Materials Engineering, Iwate University, Morioka 020-8551, Japan}
\author{Shigeki Nimori}
\affiliation{National Institute for Materials Science, Tsukuba 305-0047, Japan}
\author{Ramanathaan Suryanarayanan}
\affiliation{Laboratoire de Physico-Chimie de l'Etat Solide, CNRS, UMR8182, Universit\'{e} Paris-Sud, 91405 Orsay, France}


\date{\today}

\begin{abstract}
We investigated slow relaxations of the magnetostriction and residual magnetostriction of the phase-separated system \EGSMO, 
in which the metamagnetic transition from a paramagnetic insulating state to a ferromagnetic metallic state is accompanied by a lattice shrinkage.
The relaxations are well fitted by a stretched exponential function, suggesting the strong frustraction between the double exchange interaction and Jahn-Teller effect.
We have revealed that the Gd substitution suppresses the frozen phase-separated phase at low temperatures and stabilizes the paramagnetic insulating state in the dynamic phase-separated phase at intermediate temperatures.
The former origin would be the randomness effect and the latter would be the suppression of the double exchange interaction.
\end{abstract}


\maketitle

\section{Introduction}
Perovskite manganites fascinate many researchers with curious phenomena which are related to electron interactions: for example, colossal magnetoresistance or charge/orbital ordering.
In \RAMO~($R$ = Y, rare earth; $A$ = Ca, Sr, Ba, Pb), in which both Mn$^{3+}$ sites and Mn$^{4+}$ sites exist, the double exchange interaction competes with the Jahn-Teller effect.
Whereas the double exchange interaction causes itinerancy and lattice homogeneity, 
the Jahn-Teller effect deforms Mn$^{3+}$O$_6$ octahedra to reduce the energy level of the occupied $e_g$ orbital and localize $e_g$ electrons.
For spins, the ferromagnetic (FM) double exchange interaction competes with the antiferromagnetic (AFM) superexchange interaction.
Phase separation sometimes happens in such frustrated systems.

\LPCMO~which is an example of the phase-separated (PS) systems exhibits a magnetic-field-induced transition from an AFM charge-ordered insulating (COI) state to a FM metallic state
~\cite{Uehara1999, Podzorov2001, Ghivelder2004, Ghivelder2005, Sharma2005, Sacanell2006}.
Interestingly, as the phase diagram is shown in Ref.~\citenum{Ghivelder2005}, the temperature dependence of the transition field $H_\mathrm{c}$ is of reentrant 
and a dynamic PS phase exists below $H_\mathrm{c}$ at intermediate temperatures.
The remarkable growth of magnetization with time indicates that the system is not in equilibrium and FM clusters continuously extend in the AFM matrix.
Similar relaxation is also observed in (La$_{0.5}$Nd$_{0.5}$)$_{1.2}$Sr$_{1.8}$Mn$_2$O$_7$~\cite{Liao2006}.

Another PS system is \LPSMO, in which FM metallic (FMM) clusters are embedded in the paramagnetic insulating (PMI) matrix
~\cite{Prellier1999, Ogasawara2000, Apostu2001, Gordon2001, Matsukawa2002, Matsukawa2003, Wang2003, Matsukawa2004, Matsukawa2005, Tokunaga2005, Matsukawa2007PRB, Matsukawa2007PRL, Taniguchi2015}.
\LPSMO~also exhibits reentrant $H_\mathrm{c}(T)$.
Notably, residual effect and slow relaxation are observed in the magnetic, transport, thermodynamic, and lattice properties:
when an applied field above $H_\mathrm{c}$ is reduced to 0~T, metastable FMM state remains in much area for long time.
Surprisingly, the relaxation time estimated from the fitting by a stretched exponential function reaches about 1 day at 20~K, for example~\cite{Matsukawa2004}.
Such a long relaxation suggests a strong frustration between competing interactions.

We are focusing on a PS system \EGSMO, which exhibits a magnetic-field-induced PMI-FMM transition with reentrant $H_\mathrm{c}(T)$ 
and whose ground state consists of FMM clusters, COI clusters and the PMI matrix
~\cite{Sundaresan1997, Nakamura2004JMMM, Nakamura2004JPSJ, Wang2007, Inomata2012JMMM, Inomata2012JPCS, Nagaraja2016}.
Previous studies indicate that the lattice degree of freedom couples with the spin degree of freedom in \EGSMO;
the magnetostriction is known to exhibit a step-like decrease at the same field as the magnetization jumps, 
and the change of the lattice constant at $H_\mathrm{c}$ is actually confirmed by X-ray powder diffraction measurement~\cite{Inomata2012JMMM, Inomata2012JPCS}.

In this study, in order to clarify the properties of the phase separation of \EGSMO, we have investigated the relaxation of the magnetostriction.
By analogy with \LPCMO~or \LPSMO, we expected that \EGSMO~has a dynamic PS phase, 
in which FMM clusters gradually grow after zero field cooling and in contrast the PMI matrix slowly recovers after switching off a field above $H_\mathrm{c}$.
We have observed non-equilibrium behavior as expected, and revealed that the dynamics is described by a stretched exponential function.
On the Gd substitution effect, we have found that the dynamic PS phase is extended to lower temperatures.

\section{Experimental}
Polycrystalline samples of \EGSMO~for $x$ of 0 and 0.1 were prepared by a solid-state reaction method.
The stoichiometric mixtures of Eu$_2$O$_3$, Gd$_2$O$_3$, SrCO$_3$ and Mn$_3$O$_4$ powders were calcined in air at 1000$^\circ$C for 24~h and 1250$^\circ$C for 48~h.
The products were ground and pressed into disk-like pellets.
The pellets were sintered at 1350$^\circ$C for 36~h.
X-ray diffraction data revealed that all samples consist of a single phase with orthorhombic structures ($Pbnm$).
The lattice parameters of the parent sample ($x = 0$) are $a = 5.4424$~\AA, $b = 5.4329$~\AA, and $c = 7.664$~\AA, which is in fair agreement with a previous work~\cite{Sundaresan1997}.

As the main topic of this study, we measured the time dependence of the magnetostriction in order to reveal the transfer from a metastable state to a stable state. 
For comparison, the time dependence of the magnetization of \EGSMO~is also measured.
In the magnetostriction measurement, we measured the resistance of a strain gauge attached on a sample, 
and convert the change of the resistance to the change of the sample length by using a gauge factor $G$:
$\frac{\Delta L}{L_0} = \frac{1}{G} \frac{\Delta R}{R_0}$.
In each measurement cycle, the initial value of the sample length $L_0$ and that of the strain-gauge resistance $R_0$ are defined.
$\Delta L$ is the change of the sample length from $L_0$ and $\Delta R$ is the change of the strain-gauge resistance from $R_0$.
The type of the strain gauges is KFL-05-120-C1-11 (Kyowa Electronic Instruments).
We measured the magnetostriction perpendicular to an applied field as well as that parallel to the field.
The former and the latter were measured by using the Physical Properties Measurement System (Quantum Design) and the Magnetic Properties Measurement System (MPMS, Quantum Design), respectively.
As a reference sample for subtracting a background, the magnetostriction of a quartz glass was measured at the same time as the sample measurement.
Magnetization measurements were also performed with MPMS.

\section{Results}
\begin{figure}[htb]
\begin{center}
\includegraphics[width=3.5in]{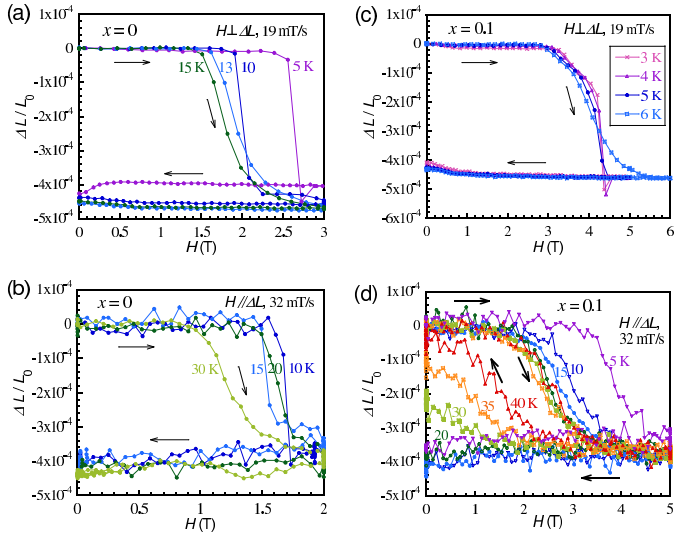}
\end{center}
\caption
{Isothermal magnetostriction of \EGSMO~after zero field cooling, which is normalized by the sample length at the virgin state. (a, b) $x=0$, (c, d) $x=0.1$. 
Every curve exhibits field-induced lattice shrinkage reflecting the transition to the FMM phase. 
}
\label{L-H}
\end{figure}

First, as shown in Fig.~\ref{L-H}, we performed magnetostriction measurements under field cycles at several temperatures below 40~K after zero field cooling.
Both the parent sample and the 10\% Gd-substituted sample exhibit negative magnetostriction, reflecting the transition from the PS or PMI phase to the FMM phase.
Corresponding to this change of the magnetostriction, in a previous study~\cite{Inomata2012JMMM}, the change of lattice constant by field is actually confirmed from X-ray diffraction measurement.
The transition field of the substituted sample is about twice as high as that of the parent sample, 
which is understood by considering the variance $\sigma^2$ of the inoic radii of $A$ site in a perovskite $AB$O$_3$~\cite{Inomata2012JMMM}.
$\sigma^2 = \sum (x_i r_i^2 -r_\mathrm{A}^2)$, in which $x_i$, $r_i$, and $r_\mathrm{A}$ are the fractional occupancies, the effective ion radii, and the average ion radius of A-site ions (Eu$^{3+}$, Gd$^{3+}$ and Sr$^{2+}$),
is estimated to be 8.66$\times$10$^{-3}$ \AA$^2$~for the parent compound and 8.79$\times$10$^{-3}$ \AA$^2$~for the substituted compound.
The higher magnitude of disorder in the substituted sample, which is indicated by the larger variance, 
is expected to suppress the double exchange interaction, enhance the energy level of the FMM state and prevent the transition to the FMM phase.
What should be noted is the remarkable residual effect: once the system transfers to the FMM phase, finite $\Delta L$ remains even after field is switched off.
This residual effect suggests that high potential barrier exists between the stable PMI state and the metastable FMM state because of some strong frustration 
and it prevents immediate transfer from the FMM state to the PMI state.
These results are consistent with a previous report~\cite{Inomata2012JMMM}.

What we newly revealed on the $L(H)$ behavior is field-direction dependence and detailed temperature dependence.
Focusing on the field-direction effect, the magnetostriction value and the transition field do not exhibit significant difference 
between the cases in which the measured sample length is parallel and perpendicular to the field.
In the temperature range of our measurement, transition field is lowered by increasing temperature.
It is because thermal energy helps the transfer from the PMI state to the FMM state beyond the potential barrier.

\begin{figure}[htb]
\begin{center}
\includegraphics[width=3.5in]{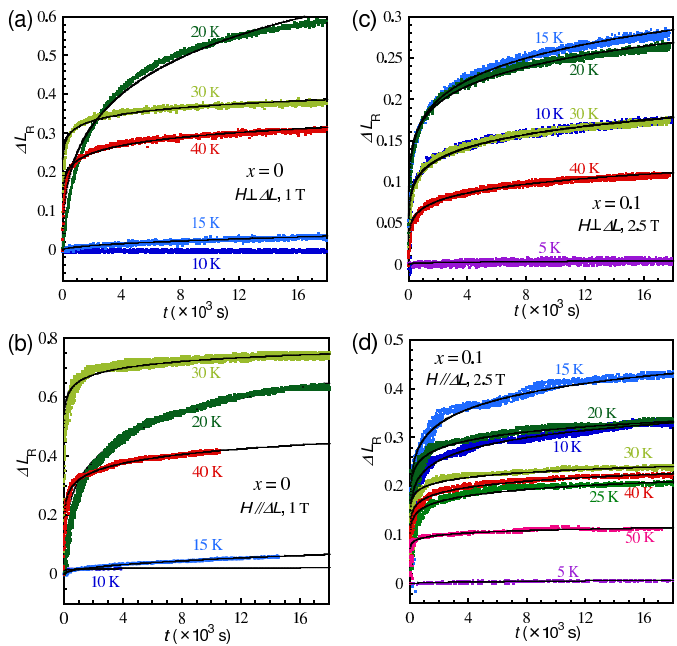}
\end{center}
\caption
{Time evolution of the relaxation rate of the magnetostriction in \EGSMO. 
After zero field cooling, the measurements were performed at (a, b) 1~T for $x$ = 0 and (c, d) 2.5~T for $x$ = 0.1. 
The increase of this \dLR~means the transfer from the PMI state to the FMM state. 
As drawn by black curves, each result is fitted by a stretched exponential function: $1 - \exp(-(t/\tau)^{\beta})$.
}
\label{L-t}
\end{figure}

Next, aiming to reveal how the volume fraction of the FMM clusters in a PS phase expands with time in \EGSMO, we measured the time evolution of the magnetostriction.
As the phase diagram is shown in Ref.~\citenum{Inomata2012JMMM}, the FMM transition field $H_\mathrm{c}$~of \EGSMO~exhibits a reentrant temperature dependence.
By analogy with the electronic phase diagram of a similar PS system \LPCMO, below $H_\mathrm{c}^\mathrm{min}$ which is the minimum value of $H_\mathrm{c}(T)$,
\EGSMO~is expected to transfer from a homogeneous PMI phase to a frozen PS phase through a dynamic PS phase on cooling.
Therefore, focusing on the possibility of these phase transitions, we measured the time dependence of the magnetostriction at $H_\mathrm{c}^\mathrm{min}$ between 5 and 50~K after zero field cooling.
Here, $H_\mathrm{c}^\mathrm{min}$ is 1~T for the parent compound and 2.5~T for the 10\%-Gd-substituted compound. 
In Fig.~\ref{L-t}, the growth of the FMM area with time is presented. 
Instead of the raw value of the magnetostriction, we introduce the relaxation rate \dLR$(t)$ and use it for the vertical axis:
\begin{equation}
\Delta L_\mathrm{R}(t) \equiv \frac{\Delta L(t) - \Delta L(0)}{\Delta L(\infty) - \Delta L(0)}.
\label{eq1}
\end{equation}
Here, $t = 0$ is the moment at which the field reaches $H_\mathrm{c}^\mathrm{min}$.
Thus \dLR~is expected to correspond to the increasing volume fraction of the FMM state.
To estimate \dLR, we have assumed that $\Delta L(\infty)$ is $\Delta L$ at $H_\mathrm{c}^\mathrm{min}$ during the field-decreasing process of the measurement shown in Fig.~\ref{L-H}.
We have clarified that the fastest evolution occurs in 20-30~K for the parent compound and at about 15~K for the 10\% Gd-substituted compound.
These results that the extension of the FMM area is fastest at an intermediate temperature is consistent 
with the scenario that the dynamic PS phase exists at a temperature range between the homogeneous PMI phase and the frozen PS phase.
Remarkable direction dependence is not observed.

\begin{figure}[htb]
\begin{center}
\includegraphics[width=3.5in]{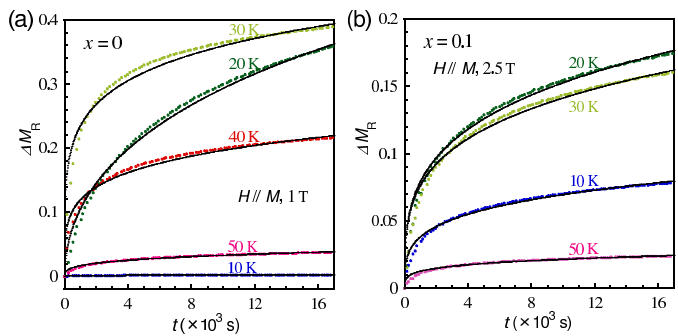}
\end{center}
\caption
{Time evolution of the relaxation rate of the magnetization in \EGSMO. 
After zero field cooling, the measurements were performed at (a) 1~T for $x$ = 0 and (b) 2.5~T for $x$ = 0.1. 
The increase of this \dMR~means the transfer from the PM state to the FMM state. 
As drawn by black curves, each result is fitted by a stretched exponential function: $1 - \exp(-(t/\tau)^{\beta})$.
}
\label{M-t}
\end{figure}

For comparison, we also measured the time dependence of the magnetization in the same condition.
\dMR~introduced in Fig.~\ref{M-t} is defined in the same manner as \dLR~in order to reflect the increasing volume fraction of the FMM state:
\dMR $ \equiv (\Delta M(t) - \Delta M(0)) / (\Delta M(\infty) - \Delta M(0))$.
We found that the time evolution of the FMM volume fraction estimated from magnetization exhibits a similar tendency to that estimated from magnetostriction for both the compounds.

\begin{figure}[htb]
\begin{center}
\includegraphics[width=3.5in]{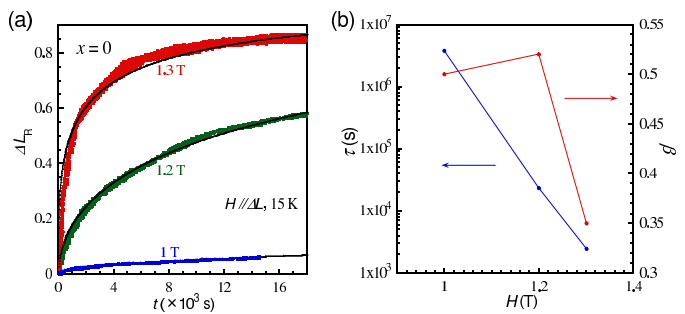}
\end{center}
\caption
{(a) Field effect on the time evolutions of the relaxation rate of the magnetostriction in the parent compound \ESMO. 
In order to detect the transfer from the PMI state to the FMM state,
magnetostriction along the field direction is measured at 15~K after zero field cooling. 
As drawn by black curves, each result is fitted by a stretched exponential function: $1 - \exp(-(t/\tau)^{\beta})$. 
(b) Field dependence of $\tau$ and $\beta$ which are the parameters of the fitting curves shown in (a).
}
\label{L-t_H-dif}
\end{figure}

For the parent sample, we also investigated the field effect on the evolution of the FMM state by measuring the magnetostriction at 15~K.
As shown in Fig.~\ref{L-t_H-dif}, \dLR (14400~s = 4~h) which is less than 0.1 under 1~T is drastically enhanced to more than 0.8 under 1.5~T.
The result indicates that increasing magnetic field changes the electronic phase of \ESMO~from the frozen PS phase to the dynamic PS phase by lowering the energy level of the FMM state.

\begin{figure}[htb]
\begin{center}
\includegraphics[width=3.5in]{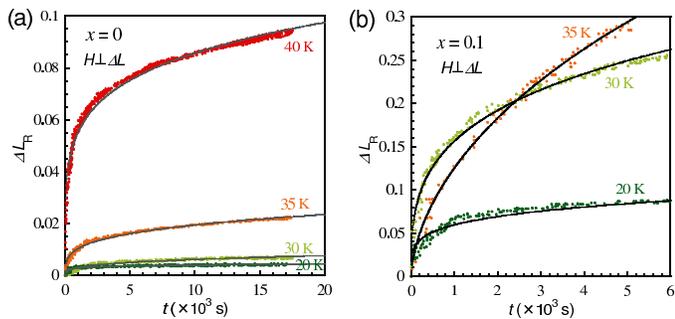}
\end{center}
\caption
{Time evolution of the relaxation rate of the residual magnetostriction in \EGSMO. 
Before the measurements at 0~T, samples were cooled in zero field and once experienced 5~T. 
(a) $x$ = 0, (b) $x$ = 0.1. The increase of this \dLR~ means the transfer from the FMM state to the PMI state. 
As drawn by black curves, each result is fitted by a stretched exponential function: $1 - \exp(-(t/\tau)^{\beta})$.
}
\label{Residual_L-t}
\end{figure}

On the other hand, in order to reveal the evolution of the PMI state, we measured the time dependence of the residual magnetostriction
after the sample was cooled in zero field and then field of 5~T is once applied and switched off.
The measurements were performed above 20~K, above which the dynamic PS phase or the homogeneous PMI phase is expected from the time evolution of the FMM volume fraction at $H_\mathrm{c}^\mathrm{min}$.
In Fig.~\ref{Residual_L-t}, the growth of the PMI area with time is presented.
Here, \dLR$(t)$ is estimated by defining $t = 0$ as the moment at which the applied field is switched off.
$\Delta L(\infty)$ corresponds to $\Delta L$ at the initial state in the field-sweep measurement shown in Fig.~\ref{L-H}.
The evolution of the PMI volume fraction becomes slower on cooling.
This behavior might suggest that the transfer from the metastable FMM state to the stable PMI state is of thermally-activated type.

\section{Discussion}
As shown in Figs.~\ref{L-t}-\ref{L-t_H-dif}, we have revealed that the time evolution of \dLR~and \dMR~is well fitted by a stretched exponential function $f(t)$:
\begin{equation}
f(t) = 1 - \exp(-(t/\tau)^{\beta}) \phantom{aaa}.
\end{equation}
Since the stretched exponential function often describes phenomena in frustrated systems, the relaxation of this system is also expected to be governed by some frustration.
In the lattice of \EGSMO, Jahn-Teller orbital-lattice interaction and double-exchange interaction will compete. Jahn-Teller effect is caused by the $e_g$ electron of \Mn3+ and deforms Mn$^{3+}$O$_6$ octahedra. 
Since \Mn3+ and Mn$^{4+}$ are mixed in \EGSMO~and Jahn-Teller distortion does not occur in Mn$^{4+}$O$_6$ octahedra, Jahn-Teller interaction causes inhomogeneous lattice. 
In contrast, double exchange interaction between \Mn3+ and Mn$^{4+}$ favors homogeneous lattice because of the itinerancy of $e_g$ electrons. 
Therefore, the slow relaxations of the magnetostriction clarified in Fig.~\ref{L-t} and \ref{Residual_L-t} suggest the strong frustration between Jahn-Teller and double exchange interactions. 
Among the spins in \EGSMO, FM double exchange interaction and AFM superexchange interaction compete. 
Thus, the slow relaxations of the magnetization in Fig.~\ref{M-t} are expected to reflect the strong magnetic frustration between double exchange and superexchange interactions.

\begin{figure}[htb]
\begin{center}
\includegraphics[width=3.5in]{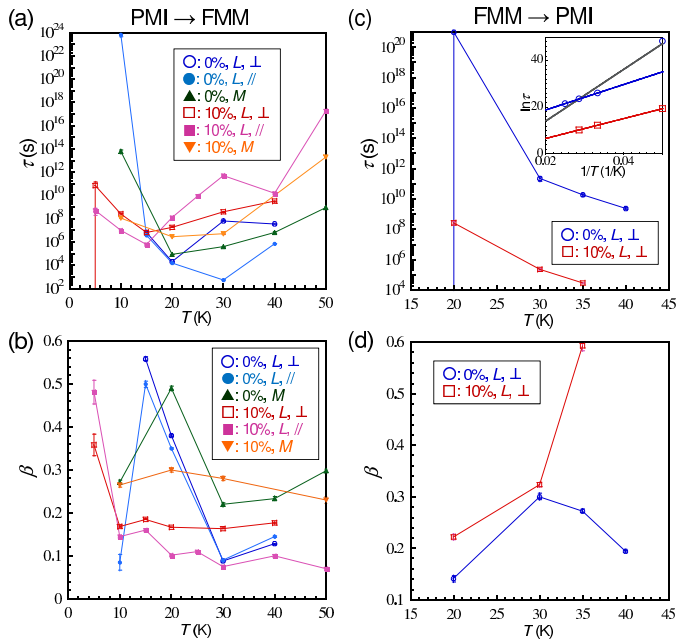}
\end{center}
\caption
{Comparison of the temperature dependence of (a, c) $\tau$ and (b, d) $\beta$ 
between different Gd compositions (0\% and 10\%) or different physical quantities (magnetostriction and magnetization). 
$\tau$ and $\beta$ of each fitting curve presented in Figs. \ref{L-t}-\ref{M-t} are plotted with error bars.
Described phenomena are slow transfer (a, b) from the PMI state to the FMM state and (c, d) from the FMM state to the PMI state.
The inset of (c) presents analyses of $\tau$ based on thermally-activated function: $\tau(T) = \tau_0 \exp(\Delta/k_\mathrm{B} T)$.
Lines are fitting curves.
For the parent compound, two fitting results are presented: $\tau$(20~K) with an extremely large error is excluded from one fitting, while it is included in the other fitting.
}
\label{tau_beta}
\end{figure}

The parameters of the fitting by a stretched exponential function, $\tau$ and $\beta$ are summarized in Figs.~\ref{L-t_H-dif}(b) and \ref{tau_beta}.
Focusing on the transfer to the FMM state shown in Fig.~\ref{tau_beta}(a) and (b), 
we have found that the order of $\tau$ estimated from the magnetostriction is similar to that from the magnetization both in the parent compound and in the Gd-substituted compound.
The result indicates that the lattice dynamics has a time scale similar to that of the spin dynamics in \EGSMO.
We have clarified that $\tau$ exhibits a minimum $\tau_\mathrm{min}$ at about 20~K for the parent compound and at about 15~K for the Gd-substituted compound with increasing temperature,
and $\beta$ of the parent compound (Fig.~\ref{tau_beta}(b)) exhibits a maximum at about 15~K while that of the substituted compound decreases monotonically.
The existence of $\tau_\mathrm{min}$ suggests that the dynamic PS phase exists around these characteristic temperatures $T^*$s, 
and the lowering $T^*$ indicates that the frozen PS phase is suppressed by the Gd substitution.
We have revealed that $\tau$ below $T^*$ is suppressed by more than 6 orders of magnitude by the substitution, which also indicates the suppression of the frozen phase.
This non-equilibrium tendency would be caused by the randomness effect introduced by the substitution.
In contrast, $\tau$ above $T^*$ increases by about 3 orders of magnitude under the substitution.
This behavior suggests that the double exchange interaction is suppressed by the enhanced disorder and the energy difference between the FMM state and the PMI state is increased.

On the transfer to the PMI state shown in Fig.~\ref{tau_beta}(c) and (d), we have clarified that $\tau$ is suppressed by 6 orders of magnitude by the Gd substitution.
The result is consistent with the enhancement of $\tau$ above $T^*$ in the transfer to the FMM state.
Importantly, we have revealed that $\tau(T)$ is described by a thermally-activated function: $\tau_0 \exp(\Delta/k_\mathrm{B} T)$.
The fitting curves are shown in the inset of Fig.~\ref{tau_beta}(c), and the thermal activation energy $\Delta$ of the substituted compound is estimated to be 37~meV.
For the parent compound, since the error of $\tau$(20~K) is extremely large, 
we ignored $\tau$(20~K) in a fitting (the blue line in the inset of Fig.~\ref{tau_beta}(c)) and obtained $\Delta$ of 47~meV.
The analysis quantitatively indicates that the Gd substitution lowers the potential barrier of the transfer from the FMM state to the PMI state.
Even if $\Delta$ of the parent compound is estimated including $\tau$(20~K) (the gray line in the inset of Fig.~\ref{tau_beta}(c)), the tendency of the substitution effect does not change.

\section{Summary}
In order to reveal non-equilibrium phenomena in PS systems, we measured the magnetostriction and residual magnetostriction of \EGSMO~($x$ = 0, 0.1);
\EGSMO~exhibits a magnetic-field-induced PMI-FMM transition which couples to a lattice shrinkage, and its ground state is a PS phase in which FMM clusters and COI clusters are embedded in a PMI matrix.
By analogy with a similar PS system \LPCMO, we expected that a dynamic PS phase exists below $H_\mathrm{c}$~between a homogeneous PMI phase at higher temperatures and a frozen PS phase at lower temperatures.
As expected, we have found continuous shrinkage of the sample length with time, which indicates the extension of the FMM clusters.

Notably, the time dependences are well fitted by stretched exponential functions, which suggests the existence of a strong frustration in \EGSMO.
As the origin of the frustration, we point out the competition between the double exchange interaction and the Jahn-Teller effect in the lattice, 
and the competition between the double exchange and superexchange interactions among spins.
From the comparison of the fitting parameter $\tau$ between magnetostriction and magnetization, we have revealed that the lattice dynamics is governed by a similar time scale to the spin dynamics.

For the Gd substitution effect, we have clarified that the frozen PS phase is suppressed, which is expected to be caused by the randomness effect.
We have also found that $\tau$ of the transfer to the FMM state is enhanced by about 3 orders of magnitude and that to the PMI state is suppressed by 6 orders of magnitude in the dynamic PS phase.
Moreover, we have revealed that $\tau(T)$ of the transfer to the PMI state is described by a thermally-activated function 
and the thermal activation energy $\Delta$ is estimated to be 47~meV for the parent compound and 37~meV for the substituted compound.
These results indicate that the Gd substitution stabilizes the PMI state. The origin would be the suppression of the double exchange interaction.

\begin{acknowledgments}
We thank M. Nakamura for his support in the measurements.
This work was supported by a Grant-in-Aid for Scientific Research from Japan Society of the Promotion of Science.
\end{acknowledgments}

\bibliography{string,Mn113,Mn327,([EuGd]Sr)MnO3_others}

\begin{thebibliography}{27}
\expandafter\ifx\csname natexlab\endcsname\relax\def\natexlab#1{#1}\fi
\expandafter\ifx\csname bibnamefont\endcsname\relax
  \def\bibnamefont#1{#1}\fi
\expandafter\ifx\csname bibfnamefont\endcsname\relax
  \def\bibfnamefont#1{#1}\fi
\expandafter\ifx\csname citenamefont\endcsname\relax
  \def\citenamefont#1{#1}\fi
\expandafter\ifx\csname url\endcsname\relax
  \def\url#1{\texttt{#1}}\fi
\expandafter\ifx\csname urlprefix\endcsname\relax\def\urlprefix{URL }\fi
\providecommand{\bibinfo}[2]{#2}
\providecommand{\eprint}[2][]{\url{#2}}

\bibitem[{\citenamefont{Uehara et~al.}(1999)\citenamefont{Uehara, Mori, Chen,
  and Cheong}}]{Uehara1999}
\bibinfo{author}{\bibfnamefont{M.}~\bibnamefont{Uehara}},
  \bibinfo{author}{\bibfnamefont{S.}~\bibnamefont{Mori}},
  \bibinfo{author}{\bibfnamefont{C.~H.} \bibnamefont{Chen}}, \bibnamefont{and}
  \bibinfo{author}{\bibfnamefont{S.-W.} \bibnamefont{Cheong}},
  \bibinfo{journal}{Nature} \textbf{\bibinfo{volume}{399}},
  \bibinfo{pages}{560} (\bibinfo{year}{1999}).

\bibitem[{\citenamefont{Podzorov et~al.}(2001)\citenamefont{Podzorov, Kim,
  Kiryukhin, Gershenson, and Cheong}}]{Podzorov2001}
\bibinfo{author}{\bibfnamefont{V.}~\bibnamefont{Podzorov}},
  \bibinfo{author}{\bibfnamefont{B.~G.} \bibnamefont{Kim}},
  \bibinfo{author}{\bibfnamefont{V.}~\bibnamefont{Kiryukhin}},
  \bibinfo{author}{\bibfnamefont{M.~E.} \bibnamefont{Gershenson}},
  \bibnamefont{and} \bibinfo{author}{\bibfnamefont{S.-W.}
  \bibnamefont{Cheong}}, \bibinfo{journal}{Phys. Rev. B}
  \textbf{\bibinfo{volume}{64}}, \bibinfo{pages}{R140406}
  (\bibinfo{year}{2001}).

\bibitem[{\citenamefont{Ghivelder et~al.}(2004)\citenamefont{Ghivelder,
  Freitas, das Virgens, Continentino, Martinho, Granja, Quintero, Leyva, Levy,
  and Parisi}}]{Ghivelder2004}
\bibinfo{author}{\bibfnamefont{L.}~\bibnamefont{Ghivelder}},
  \bibinfo{author}{\bibfnamefont{R.~S.} \bibnamefont{Freitas}},
  \bibinfo{author}{\bibfnamefont{M.~G.} \bibnamefont{das Virgens}},
  \bibinfo{author}{\bibfnamefont{M.~A.} \bibnamefont{Continentino}},
  \bibinfo{author}{\bibfnamefont{H.}~\bibnamefont{Martinho}},
  \bibinfo{author}{\bibfnamefont{L.}~\bibnamefont{Granja}},
  \bibinfo{author}{\bibfnamefont{M.}~\bibnamefont{Quintero}},
  \bibinfo{author}{\bibfnamefont{G.}~\bibnamefont{Leyva}},
  \bibinfo{author}{\bibfnamefont{P.}~\bibnamefont{Levy}}, \bibnamefont{and}
  \bibinfo{author}{\bibfnamefont{F.}~\bibnamefont{Parisi}},
  \bibinfo{journal}{Phys. Rev. B} \textbf{\bibinfo{volume}{69}},
  \bibinfo{pages}{214414} (\bibinfo{year}{2004}).

\bibitem[{\citenamefont{Ghivelder and Parisi}(2005)}]{Ghivelder2005}
\bibinfo{author}{\bibfnamefont{L.}~\bibnamefont{Ghivelder}} \bibnamefont{and}
  \bibinfo{author}{\bibfnamefont{F.}~\bibnamefont{Parisi}},
  \bibinfo{journal}{Phys. Rev. B} \textbf{\bibinfo{volume}{71}},
  \bibinfo{pages}{184425} (\bibinfo{year}{2005}).

\bibitem[{\citenamefont{Sharma et~al.}(2005)\citenamefont{Sharma, Kim, Koo,
  Guha, and Cheong}}]{Sharma2005}
\bibinfo{author}{\bibfnamefont{P.~A.} \bibnamefont{Sharma}},
  \bibinfo{author}{\bibfnamefont{S.~B.} \bibnamefont{Kim}},
  \bibinfo{author}{\bibfnamefont{T.~Y.} \bibnamefont{Koo}},
  \bibinfo{author}{\bibfnamefont{S.}~\bibnamefont{Guha}}, \bibnamefont{and}
  \bibinfo{author}{\bibfnamefont{S.-W.} \bibnamefont{Cheong}},
  \bibinfo{journal}{Phys. Rev. B} \textbf{\bibinfo{volume}{71}},
  \bibinfo{pages}{224416} (\bibinfo{year}{2005}).

\bibitem[{\citenamefont{Sacanell et~al.}(2006)\citenamefont{Sacanell, Parisi,
  Campoy, and Ghivelder}}]{Sacanell2006}
\bibinfo{author}{\bibfnamefont{J.}~\bibnamefont{Sacanell}},
  \bibinfo{author}{\bibfnamefont{F.}~\bibnamefont{Parisi}},
  \bibinfo{author}{\bibfnamefont{J.~C.~P.} \bibnamefont{Campoy}},
  \bibnamefont{and}
  \bibinfo{author}{\bibfnamefont{L.}~\bibnamefont{Ghivelder}},
  \bibinfo{journal}{Phys. Rev. B} \textbf{\bibinfo{volume}{73}},
  \bibinfo{pages}{014403} (\bibinfo{year}{2006}).

\bibitem[{\citenamefont{qian Liao et~al.}(2006)\citenamefont{qian Liao, Sun,
  fu~Yang, an~Li, and hua Cheng}}]{Liao2006}
\bibinfo{author}{\bibfnamefont{D.}~\bibnamefont{qian Liao}},
  \bibinfo{author}{\bibfnamefont{Y.}~\bibnamefont{Sun}},
  \bibinfo{author}{\bibfnamefont{R.}~\bibnamefont{fu~Yang}},
  \bibinfo{author}{\bibfnamefont{Q.}~\bibnamefont{an~Li}}, \bibnamefont{and}
  \bibinfo{author}{\bibfnamefont{Z.}~\bibnamefont{hua Cheng}},
  \bibinfo{journal}{Phys. Rev. B} \textbf{\bibinfo{volume}{74}},
  \bibinfo{pages}{174434} (\bibinfo{year}{2006}).

\bibitem[{\citenamefont{Prellier et~al.}(1999)\citenamefont{Prellier,
  Suryanarayanan, Dhalenne, Berthon, Renard, Dupas, and
  Revcolevschi}}]{Prellier1999}
\bibinfo{author}{\bibfnamefont{W.}~\bibnamefont{Prellier}},
  \bibinfo{author}{\bibfnamefont{R.}~\bibnamefont{Suryanarayanan}},
  \bibinfo{author}{\bibfnamefont{G.}~\bibnamefont{Dhalenne}},
  \bibinfo{author}{\bibfnamefont{J.}~\bibnamefont{Berthon}},
  \bibinfo{author}{\bibfnamefont{J.-P.} \bibnamefont{Renard}},
  \bibinfo{author}{\bibfnamefont{C.}~\bibnamefont{Dupas}}, \bibnamefont{and}
  \bibinfo{author}{\bibfnamefont{A.}~\bibnamefont{Revcolevschi}},
  \bibinfo{journal}{Physica B} \textbf{\bibinfo{volume}{259}},
  \bibinfo{pages}{833} (\bibinfo{year}{1999}).

\bibitem[{\citenamefont{Ogasawara et~al.}(2000)\citenamefont{Ogasawara,
  Matsukawa, Hatakeyama, Yoshizawa, Apostu, Suryanarayanan, Dhalenne,
  Revcolevschi, Itoh, and Kobayashi}}]{Ogasawara2000}
\bibinfo{author}{\bibfnamefont{H.}~\bibnamefont{Ogasawara}},
  \bibinfo{author}{\bibfnamefont{M.}~\bibnamefont{Matsukawa}},
  \bibinfo{author}{\bibfnamefont{S.}~\bibnamefont{Hatakeyama}},
  \bibinfo{author}{\bibfnamefont{M.}~\bibnamefont{Yoshizawa}},
  \bibinfo{author}{\bibfnamefont{M.}~\bibnamefont{Apostu}},
  \bibinfo{author}{\bibfnamefont{R.}~\bibnamefont{Suryanarayanan}},
  \bibinfo{author}{\bibfnamefont{G.}~\bibnamefont{Dhalenne}},
  \bibinfo{author}{\bibfnamefont{A.}~\bibnamefont{Revcolevschi}},
  \bibinfo{author}{\bibfnamefont{K.}~\bibnamefont{Itoh}}, \bibnamefont{and}
  \bibinfo{author}{\bibfnamefont{N.}~\bibnamefont{Kobayashi}},
  \bibinfo{journal}{J. Phys. Soc. Jpn.} \textbf{\bibinfo{volume}{69}},
  \bibinfo{pages}{1274} (\bibinfo{year}{2000}).

\bibitem[{\citenamefont{Apostu et~al.}(2001)\citenamefont{Apostu,
  Suryanarayanan, Revcolevschi, Ogasawara, Matsukawa, Yoshizawa, and
  Kobayashi}}]{Apostu2001}
\bibinfo{author}{\bibfnamefont{M.}~\bibnamefont{Apostu}},
  \bibinfo{author}{\bibfnamefont{R.}~\bibnamefont{Suryanarayanan}},
  \bibinfo{author}{\bibfnamefont{A.}~\bibnamefont{Revcolevschi}},
  \bibinfo{author}{\bibfnamefont{H.}~\bibnamefont{Ogasawara}},
  \bibinfo{author}{\bibfnamefont{M.}~\bibnamefont{Matsukawa}},
  \bibinfo{author}{\bibfnamefont{M.}~\bibnamefont{Yoshizawa}},
  \bibnamefont{and}
  \bibinfo{author}{\bibfnamefont{N.}~\bibnamefont{Kobayashi}},
  \bibinfo{journal}{Phys. Rev. B} \textbf{\bibinfo{volume}{64}},
  \bibinfo{pages}{012407} (\bibinfo{year}{2001}).

\bibitem[{\citenamefont{Gordon et~al.}(2001)\citenamefont{Gordon, Wagner,
  Moshchalkov, Bruynseraede, Apostu, Suryanarayanan, and
  Revcolevschi}}]{Gordon2001}
\bibinfo{author}{\bibfnamefont{I.}~\bibnamefont{Gordon}},
  \bibinfo{author}{\bibfnamefont{P.}~\bibnamefont{Wagner}},
  \bibinfo{author}{\bibfnamefont{V.~V.} \bibnamefont{Moshchalkov}},
  \bibinfo{author}{\bibfnamefont{Y.}~\bibnamefont{Bruynseraede}},
  \bibinfo{author}{\bibfnamefont{M.}~\bibnamefont{Apostu}},
  \bibinfo{author}{\bibfnamefont{R.}~\bibnamefont{Suryanarayanan}},
  \bibnamefont{and}
  \bibinfo{author}{\bibfnamefont{A.}~\bibnamefont{Revcolevschi}},
  \bibinfo{journal}{Phys. Rev. B} \textbf{\bibinfo{volume}{64}},
  \bibinfo{pages}{092408} (\bibinfo{year}{2001}).

\bibitem[{\citenamefont{Matsukawa et~al.}(2002)\citenamefont{Matsukawa,
  Ogasawara, Sasaki, Yoshizawa, Apostu, Suryanarayanan, Revcolevschi, Itoh, and
  Kobayashi}}]{Matsukawa2002}
\bibinfo{author}{\bibfnamefont{M.}~\bibnamefont{Matsukawa}},
  \bibinfo{author}{\bibfnamefont{H.}~\bibnamefont{Ogasawara}},
  \bibinfo{author}{\bibfnamefont{T.}~\bibnamefont{Sasaki}},
  \bibinfo{author}{\bibfnamefont{M.}~\bibnamefont{Yoshizawa}},
  \bibinfo{author}{\bibfnamefont{M.}~\bibnamefont{Apostu}},
  \bibinfo{author}{\bibfnamefont{R.}~\bibnamefont{Suryanarayanan}},
  \bibinfo{author}{\bibfnamefont{A.}~\bibnamefont{Revcolevschi}},
  \bibinfo{author}{\bibfnamefont{K.}~\bibnamefont{Itoh}}, \bibnamefont{and}
  \bibinfo{author}{\bibfnamefont{N.}~\bibnamefont{Kobayashi}},
  \bibinfo{journal}{J. Phys. Soc. Jpn.} \textbf{\bibinfo{volume}{71}},
  \bibinfo{pages}{1475} (\bibinfo{year}{2002}).

\bibitem[{\citenamefont{Matsukawa et~al.}(2003)\citenamefont{Matsukawa, Narita,
  Nishimura, Yoshizawa, Apostu, Suryanarayanan, Revcolevschi, Itoh, and
  Kobayashi}}]{Matsukawa2003}
\bibinfo{author}{\bibfnamefont{M.}~\bibnamefont{Matsukawa}},
  \bibinfo{author}{\bibfnamefont{M.}~\bibnamefont{Narita}},
  \bibinfo{author}{\bibfnamefont{T.}~\bibnamefont{Nishimura}},
  \bibinfo{author}{\bibfnamefont{M.}~\bibnamefont{Yoshizawa}},
  \bibinfo{author}{\bibfnamefont{M.}~\bibnamefont{Apostu}},
  \bibinfo{author}{\bibfnamefont{R.}~\bibnamefont{Suryanarayanan}},
  \bibinfo{author}{\bibfnamefont{A.}~\bibnamefont{Revcolevschi}},
  \bibinfo{author}{\bibfnamefont{K.}~\bibnamefont{Itoh}}, \bibnamefont{and}
  \bibinfo{author}{\bibfnamefont{N.}~\bibnamefont{Kobayashi}},
  \bibinfo{journal}{Phys. Rev. B} \textbf{\bibinfo{volume}{67}},
  \bibinfo{pages}{104433} (\bibinfo{year}{2003}).

\bibitem[{\citenamefont{Wang et~al.}(2003)\citenamefont{Wang, Gukasov, Moussa,
  Hennion, Apostu, Suryanarayanan, and Revcolevschi}}]{Wang2003}
\bibinfo{author}{\bibfnamefont{F.}~\bibnamefont{Wang}},
  \bibinfo{author}{\bibfnamefont{A.}~\bibnamefont{Gukasov}},
  \bibinfo{author}{\bibfnamefont{F.}~\bibnamefont{Moussa}},
  \bibinfo{author}{\bibfnamefont{M.}~\bibnamefont{Hennion}},
  \bibinfo{author}{\bibfnamefont{M.}~\bibnamefont{Apostu}},
  \bibinfo{author}{\bibfnamefont{R.}~\bibnamefont{Suryanarayanan}},
  \bibnamefont{and}
  \bibinfo{author}{\bibfnamefont{A.}~\bibnamefont{Revcolevschi}},
  \bibinfo{journal}{Phys. Rev. Lett.} \textbf{\bibinfo{volume}{91}},
  \bibinfo{pages}{047204} (\bibinfo{year}{2003}).

\bibitem[{\citenamefont{Matsukawa et~al.}(2004)\citenamefont{Matsukawa, Chiba,
  Akasaka, Suryanarayanan, Apostu, Revcolevschi, Nimori, and
  Kobayashi}}]{Matsukawa2004}
\bibinfo{author}{\bibfnamefont{M.}~\bibnamefont{Matsukawa}},
  \bibinfo{author}{\bibfnamefont{M.}~\bibnamefont{Chiba}},
  \bibinfo{author}{\bibfnamefont{A.}~\bibnamefont{Akasaka}},
  \bibinfo{author}{\bibfnamefont{R.}~\bibnamefont{Suryanarayanan}},
  \bibinfo{author}{\bibfnamefont{M.}~\bibnamefont{Apostu}},
  \bibinfo{author}{\bibfnamefont{A.}~\bibnamefont{Revcolevschi}},
  \bibinfo{author}{\bibfnamefont{S.}~\bibnamefont{Nimori}}, \bibnamefont{and}
  \bibinfo{author}{\bibfnamefont{N.}~\bibnamefont{Kobayashi}},
  \bibinfo{journal}{Phys. Rev. B} \textbf{\bibinfo{volume}{70}},
  \bibinfo{pages}{132402} (\bibinfo{year}{2004}).

\bibitem[{\citenamefont{Matsukawa et~al.}(2005)\citenamefont{Matsukawa,
  Akasaka, Noto, Suryanarayanan, Nimori, Apostu, Revcolevschi, and
  Kobayashi}}]{Matsukawa2005}
\bibinfo{author}{\bibfnamefont{M.}~\bibnamefont{Matsukawa}},
  \bibinfo{author}{\bibfnamefont{K.}~\bibnamefont{Akasaka}},
  \bibinfo{author}{\bibfnamefont{H.}~\bibnamefont{Noto}},
  \bibinfo{author}{\bibfnamefont{R.}~\bibnamefont{Suryanarayanan}},
  \bibinfo{author}{\bibfnamefont{S.}~\bibnamefont{Nimori}},
  \bibinfo{author}{\bibfnamefont{M.}~\bibnamefont{Apostu}},
  \bibinfo{author}{\bibfnamefont{A.}~\bibnamefont{Revcolevschi}},
  \bibnamefont{and}
  \bibinfo{author}{\bibfnamefont{N.}~\bibnamefont{Kobayashi}},
  \bibinfo{journal}{Phys. Rev. B} \textbf{\bibinfo{volume}{72}},
  \bibinfo{pages}{064412} (\bibinfo{year}{2005}).

\bibitem[{\citenamefont{Tokunaga et~al.}(2005)\citenamefont{Tokunaga, Tokunaga,
  and Tamegai}}]{Tokunaga2005}
\bibinfo{author}{\bibfnamefont{Y.}~\bibnamefont{Tokunaga}},
  \bibinfo{author}{\bibfnamefont{M.}~\bibnamefont{Tokunaga}}, \bibnamefont{and}
  \bibinfo{author}{\bibfnamefont{T.}~\bibnamefont{Tamegai}},
  \bibinfo{journal}{Phys. Rev. B} \textbf{\bibinfo{volume}{71}},
  \bibinfo{pages}{012408} (\bibinfo{year}{2005}).

\bibitem[{\citenamefont{Matsukawa
  et~al.}(2007{\natexlab{a}})\citenamefont{Matsukawa, Tamura, Nimori,
  Suryanarayanan, Kumagai, Nakanishi, Apostu, Revcolevschi, Koyama, and
  Kobayashi}}]{Matsukawa2007PRB}
\bibinfo{author}{\bibfnamefont{M.}~\bibnamefont{Matsukawa}},
  \bibinfo{author}{\bibfnamefont{A.}~\bibnamefont{Tamura}},
  \bibinfo{author}{\bibfnamefont{S.}~\bibnamefont{Nimori}},
  \bibinfo{author}{\bibfnamefont{R.}~\bibnamefont{Suryanarayanan}},
  \bibinfo{author}{\bibfnamefont{T.}~\bibnamefont{Kumagai}},
  \bibinfo{author}{\bibfnamefont{Y.}~\bibnamefont{Nakanishi}},
  \bibinfo{author}{\bibfnamefont{M.}~\bibnamefont{Apostu}},
  \bibinfo{author}{\bibfnamefont{A.}~\bibnamefont{Revcolevschi}},
  \bibinfo{author}{\bibfnamefont{K.}~\bibnamefont{Koyama}}, \bibnamefont{and}
  \bibinfo{author}{\bibfnamefont{N.}~\bibnamefont{Kobayashi}},
  \bibinfo{journal}{Phys. Rev. B} \textbf{\bibinfo{volume}{75}},
  \bibinfo{pages}{014427} (\bibinfo{year}{2007}{\natexlab{a}}).

\bibitem[{\citenamefont{Matsukawa
  et~al.}(2007{\natexlab{b}})\citenamefont{Matsukawa, Yamato, Kumagai, Tamura,
  Suryanarayanan, Nimori, Apostu, Revcolevschi, Koyama, and
  Kobayashi}}]{Matsukawa2007PRL}
\bibinfo{author}{\bibfnamefont{M.}~\bibnamefont{Matsukawa}},
  \bibinfo{author}{\bibfnamefont{Y.}~\bibnamefont{Yamato}},
  \bibinfo{author}{\bibfnamefont{T.}~\bibnamefont{Kumagai}},
  \bibinfo{author}{\bibfnamefont{A.}~\bibnamefont{Tamura}},
  \bibinfo{author}{\bibfnamefont{R.}~\bibnamefont{Suryanarayanan}},
  \bibinfo{author}{\bibfnamefont{S.}~\bibnamefont{Nimori}},
  \bibinfo{author}{\bibfnamefont{M.}~\bibnamefont{Apostu}},
  \bibinfo{author}{\bibfnamefont{A.}~\bibnamefont{Revcolevschi}},
  \bibinfo{author}{\bibfnamefont{K.}~\bibnamefont{Koyama}}, \bibnamefont{and}
  \bibinfo{author}{\bibfnamefont{N.}~\bibnamefont{Kobayashi}},
  \bibinfo{journal}{Phys. Rev. Lett.} \textbf{\bibinfo{volume}{98}},
  \bibinfo{pages}{267204} (\bibinfo{year}{2007}{\natexlab{b}}).

\bibitem[{\citenamefont{Taniguchi et~al.}(2015)\citenamefont{Taniguchi,
  Nakamura, Konno, Matsukawa, and Suryanarayanan}}]{Taniguchi2015}
\bibinfo{author}{\bibfnamefont{H.}~\bibnamefont{Taniguchi}},
  \bibinfo{author}{\bibfnamefont{Y.}~\bibnamefont{Nakamura}},
  \bibinfo{author}{\bibfnamefont{T.}~\bibnamefont{Konno}},
  \bibinfo{author}{\bibfnamefont{M.}~\bibnamefont{Matsukawa}},
  \bibnamefont{and}
  \bibinfo{author}{\bibfnamefont{R.}~\bibnamefont{Suryanarayanan}},
  \bibinfo{journal}{Physics Procedia} \textbf{\bibinfo{volume}{75}},
  \bibinfo{pages}{468} (\bibinfo{year}{2015}).

\bibitem[{\citenamefont{Sundaresan et~al.}(1997)\citenamefont{Sundaresan,
  Maignan, and Raveau}}]{Sundaresan1997}
\bibinfo{author}{\bibfnamefont{A.}~\bibnamefont{Sundaresan}},
  \bibinfo{author}{\bibfnamefont{A.}~\bibnamefont{Maignan}}, \bibnamefont{and}
  \bibinfo{author}{\bibfnamefont{B.}~\bibnamefont{Raveau}},
  \bibinfo{journal}{Phys. Rev. B} \textbf{\bibinfo{volume}{55}},
  \bibinfo{pages}{5596} (\bibinfo{year}{1997}).

\bibitem[{\citenamefont{Nakamura
  et~al.}(2004{\natexlab{a}})\citenamefont{Nakamura, Shimomura, Ikeda,
  Mizumaki, Ohsumi, Nimori, Takeuchi, and Itoh}}]{Nakamura2004JMMM}
\bibinfo{author}{\bibfnamefont{S.}~\bibnamefont{Nakamura}},
  \bibinfo{author}{\bibfnamefont{S.}~\bibnamefont{Shimomura}},
  \bibinfo{author}{\bibfnamefont{N.}~\bibnamefont{Ikeda}},
  \bibinfo{author}{\bibfnamefont{M.}~\bibnamefont{Mizumaki}},
  \bibinfo{author}{\bibfnamefont{H.}~\bibnamefont{Ohsumi}},
  \bibinfo{author}{\bibfnamefont{S.}~\bibnamefont{Nimori}},
  \bibinfo{author}{\bibfnamefont{T.}~\bibnamefont{Takeuchi}}, \bibnamefont{and}
  \bibinfo{author}{\bibfnamefont{K.}~\bibnamefont{Itoh}}, \bibinfo{journal}{J.
  Mag. Mag. Mat.} \textbf{\bibinfo{volume}{272}}, \bibinfo{pages}{424}
  (\bibinfo{year}{2004}{\natexlab{a}}).

\bibitem[{\citenamefont{Nakamura
  et~al.}(2004{\natexlab{b}})\citenamefont{Nakamura, Shimomura, Ikeda, Nimori,
  Takeuchi, and Itoh}}]{Nakamura2004JPSJ}
\bibinfo{author}{\bibfnamefont{S.}~\bibnamefont{Nakamura}},
  \bibinfo{author}{\bibfnamefont{S.}~\bibnamefont{Shimomura}},
  \bibinfo{author}{\bibfnamefont{N.}~\bibnamefont{Ikeda}},
  \bibinfo{author}{\bibfnamefont{S.}~\bibnamefont{Nimori}},
  \bibinfo{author}{\bibfnamefont{T.}~\bibnamefont{Takeuchi}}, \bibnamefont{and}
  \bibinfo{author}{\bibfnamefont{K.}~\bibnamefont{Itoh}}, \bibinfo{journal}{J.
  Phys. Soc. Jpn.} \textbf{\bibinfo{volume}{73}}, \bibinfo{pages}{3059}
  (\bibinfo{year}{2004}{\natexlab{b}}).

\bibitem[{\citenamefont{Wang et~al.}(2007)\citenamefont{Wang, Sun, Liu, Xie,
  Wang, Zhao, Shen, and Li}}]{Wang2007}
\bibinfo{author}{\bibfnamefont{J.~Z.} \bibnamefont{Wang}},
  \bibinfo{author}{\bibfnamefont{J.~R.} \bibnamefont{Sun}},
  \bibinfo{author}{\bibfnamefont{G.~J.} \bibnamefont{Liu}},
  \bibinfo{author}{\bibfnamefont{Y.~W.} \bibnamefont{Xie}},
  \bibinfo{author}{\bibfnamefont{D.~J.} \bibnamefont{Wang}},
  \bibinfo{author}{\bibfnamefont{T.~Y.} \bibnamefont{Zhao}},
  \bibinfo{author}{\bibfnamefont{B.~G.} \bibnamefont{Shen}}, \bibnamefont{and}
  \bibinfo{author}{\bibfnamefont{X.~G.} \bibnamefont{Li}},
  \bibinfo{journal}{Phys. Rev. B} \textbf{\bibinfo{volume}{76}},
  \bibinfo{pages}{104428} (\bibinfo{year}{2007}).

\bibitem[{\citenamefont{Inomata
  et~al.}(2012{\natexlab{a}})\citenamefont{Inomata, Matsukawa, Kimura, Yamato,
  Kobayashi, Suryanarayanan, Nimori, Koyama, Takahashi, Watanabe
  et~al.}}]{Inomata2012JMMM}
\bibinfo{author}{\bibfnamefont{T.}~\bibnamefont{Inomata}},
  \bibinfo{author}{\bibfnamefont{M.}~\bibnamefont{Matsukawa}},
  \bibinfo{author}{\bibfnamefont{D.}~\bibnamefont{Kimura}},
  \bibinfo{author}{\bibfnamefont{Y.}~\bibnamefont{Yamato}},
  \bibinfo{author}{\bibfnamefont{S.}~\bibnamefont{Kobayashi}},
  \bibinfo{author}{\bibfnamefont{R.}~\bibnamefont{Suryanarayanan}},
  \bibinfo{author}{\bibfnamefont{S.}~\bibnamefont{Nimori}},
  \bibinfo{author}{\bibfnamefont{K.}~\bibnamefont{Koyama}},
  \bibinfo{author}{\bibfnamefont{K.}~\bibnamefont{Takahashi}},
  \bibinfo{author}{\bibfnamefont{K.}~\bibnamefont{Watanabe}},
  \bibnamefont{et~al.}, \bibinfo{journal}{J. Mag. Mag. Mat.}
  \textbf{\bibinfo{volume}{324}}, \bibinfo{pages}{3863}
  (\bibinfo{year}{2012}{\natexlab{a}}).

\bibitem[{\citenamefont{Inomata
  et~al.}(2012{\natexlab{b}})\citenamefont{Inomata, Nakanishi, Matsukawa,
  Kobayashi, Nimori, Suryanarayanan, and Kobayashi}}]{Inomata2012JPCS}
\bibinfo{author}{\bibfnamefont{T.}~\bibnamefont{Inomata}},
  \bibinfo{author}{\bibfnamefont{Y.}~\bibnamefont{Nakanishi}},
  \bibinfo{author}{\bibfnamefont{M.}~\bibnamefont{Matsukawa}},
  \bibinfo{author}{\bibfnamefont{S.}~\bibnamefont{Kobayashi}},
  \bibinfo{author}{\bibfnamefont{S.}~\bibnamefont{Nimori}},
  \bibinfo{author}{\bibfnamefont{R.}~\bibnamefont{Suryanarayanan}},
  \bibnamefont{and}
  \bibinfo{author}{\bibfnamefont{N.}~\bibnamefont{Kobayashi}},
  \bibinfo{journal}{J. Phys.: Conf. Ser.} \textbf{\bibinfo{volume}{400}},
  \bibinfo{pages}{032051} (\bibinfo{year}{2012}{\natexlab{b}}).

\bibitem[{\citenamefont{Nagaraja et~al.}(2016)\citenamefont{Nagaraja, Rao, and
  Okram}}]{Nagaraja2016}
\bibinfo{author}{\bibfnamefont{B.~S.} \bibnamefont{Nagaraja}},
  \bibinfo{author}{\bibfnamefont{A.}~\bibnamefont{Rao}}, \bibnamefont{and}
  \bibinfo{author}{\bibfnamefont{G.~S.} \bibnamefont{Okram}},
  \bibinfo{journal}{J. Alloys Compd.} \textbf{\bibinfo{volume}{683}},
  \bibinfo{pages}{308} (\bibinfo{year}{2016}).

\end{thebibliography}

\end{document}